%% file: main.tex
\documentclass{sig-alternate-05-2015}

\usepackage{tabularx}
\usepackage{wrapfig}
\usepackage{multirow}
\usepackage{hyperref}

\newcommand{\enquote}[1]{``#1''} 
\newcommand{\up}[1]{\ensuremath{^{\textrm{\small#1}}}} 

\usepackage{tikz}
\newcommand\copyrighttext{%
  \footnotesize \textcopyright~ACM 2016. This is the author's version of the work. It is posted here for your personal use. Not for redistribution. The definitive Version of Record was published in Proceedings of the 38th International Conference on Software Engineering Companion (ICSE'16). ACM, New York, NY, USA, 313-322, \url{https://doi.org/10.1145/2889160.2889182}.
}
\newcommand\copyrightnoticepreprint{%
    \begin{tikzpicture}[remember picture,overlay]
        \node[anchor=south,xshift=33.2em, yshift=10pt] at (current page.south) {\fbox{\parbox{\dimexpr\textwidth-\fboxsep-\fboxrule\relax}{\copyrighttext}}};
    \end{tikzpicture}%
}

\begin{document}

\setcopyright{acmcopyright}

\doi{https://doi.org/10.1145/2889160.2889182}

\isbn{978-1-4503-4205-6}

\conferenceinfo{ICSE '16}{May 14--22, 2016, Austin, TX, USA}

\acmPrice{\$15.00}

%
\conferenceinfo{ICSE '16}{May 14--22, 2016, Austin, TX, USA}
\CopyrightYear{2016} 
\crdata{978-1-4503-4205-6}  

\title{How Surveys, Tutors, and Software Help to Assess Scrum Adoption in a Classroom Software Engineering Project}
%
%
%
%
%

\numberofauthors{5} 
%
\author{
%
%
Christoph Matthies, Thomas Kowark, Keven Richly, \\ Matthias Uflacker, and Hasso Plattner\\
       \affaddr{Hasso Plattner Institute, University of Potsdam}\\
       \affaddr{August-Bebel-Str. 88}\\
       \affaddr{Potsdam, Germany}\\
       \email{\{firstname.lastname\}@hpi.de}
}
\date{30 July 1999}

\maketitle

\begin{abstract}
\input{abstract}
\end{abstract}

\copyrightnoticepreprint

%
%

\ccsdesc[500]{Software and its engineering~Agile software development}
\ccsdesc[500]{Applied computing~Collaborative learning}

%
%

%
%
\printccsdesc


\keywords{Classroom project, Scrum, multi-level assessment, collaboration analysis}

\input{intro}
\input{course_setup}
\input{surveys}
\input{tutor_scores}
\input{tool_analysis}
\input{scrumlint}
\input{learnings}
\input{related_work}
\input{conclusion}

%
\bibliographystyle{abbrv}

%
%

\end{document}

%% file: abstract.tex
Agile methods are best taught in a hands-on fashion in realistic projects.
The main challenge in doing so is to assess whether students apply the methods correctly without requiring complete supervision throughout the entire project.

This paper presents experiences from a classroom project where 38 students developed a single system using a scaled version of Scrum.
Surveys helped us to identify which elements of Scrum correlated most with student satisfaction or posed the biggest challenges.
These insights were augmented by a team of tutors, which accompanied main meetings throughout the project to provide feedback to the teams, and captured impressions of method application in practice.
Finally, we performed a post-hoc, tool-supported analysis of collaboration artifacts to detect concrete indicators for anti-patterns in Scrum adoption.

Through the combination of these techniques we were able to understand how students implemented Scrum in this course and which elements require further lecturing and tutoring in future iterations. Automated analysis of collaboration artifacts proved to be a promising addition to the development process that could potentially reduce manual efforts in future courses and allow for more concrete and targeted feedback, as well as more objective assessment.

%% file: intro.tex
\section{Introduction \& Motivation}
Classroom projects allow students to deepen their theoretical understanding of software engineering practices through first-hand experience \cite{Coppit:2005aa, Paasivaara:2013, JointTask2013}. Contrary to theoretical foundations, whose understanding can be assessed through exams, the quality of practical application is difficult to monitor \cite{Igaki2014}. If teachers do not accompany students throughout the entire project, how can they be certain that the taught methods were applied correctly?

A straightforward solution is to allow project work only when teaching staff is present. This works for smaller assignments, but does not scale to multiple teams in longer lasting projects. Also, an essential trait of agile methods is to adapt to the circumstances and find solutions that work in the given context instead of blindly adhering to a prescribed process \cite{beck2001agile}. Hence, a laboratory-like setting potentially reduces the learning experience for students \cite{Devedzic2011}. The other extreme is to focus on project outcome and discuss potential problems in retrospection sessions. This relies on students being conscientious, honest, and open about mistakes they made throughout the project. Depending on the course setup and grading criteria, students might refrain from admitting mistakes or are not necessarily aware that certain practices are detrimental to project progress.

A middle ground that combines strengths and weaknesses of both approaches is represented by regular synchronization points, where the teaching staff assesses progress and discusses process implementation with the students. Despite this more fine-grained insight into process adoption, the general problems of relying on student input and snapshot-like observation remain. So, how is it possible to unobtrusively observe the way that students employ the proposed development processes without sacrificing the freedom that agile methods have to offer?

\begin{figure}[!ht]
	\centering
	\includegraphics[width=0.9\columnwidth]{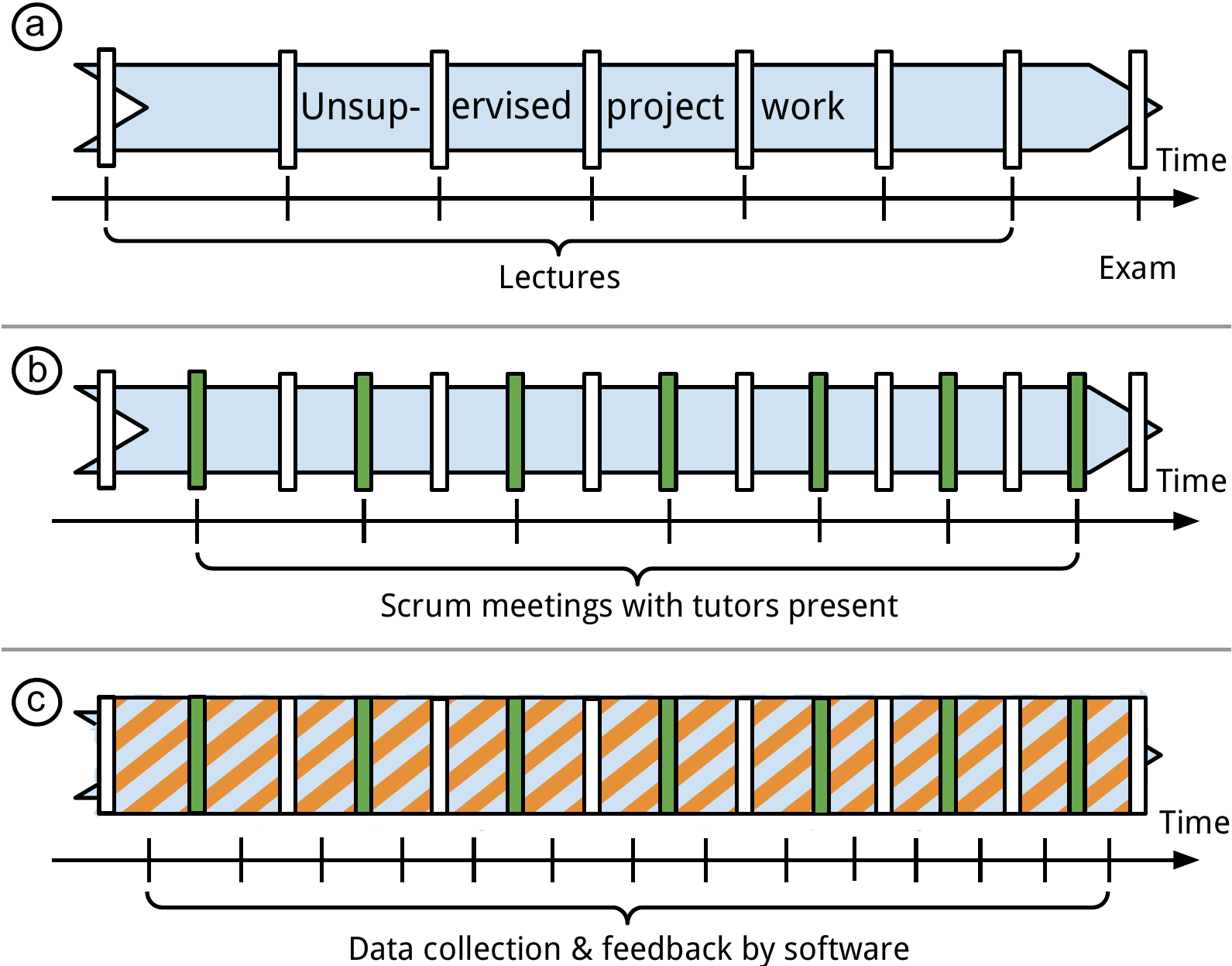}
	\caption{Giving feedback during a software engineering course. (a) Feedback opportunities in a classroom-style lecture. (b) Tutoring students during Scrum meetings. (c) Using software to provide feedback during project work.
}
	\label{fig:coverage}
\end{figure}

In this paper, we present experiences from a four month software engineering project in which 38 students jointly developed a single system (Section 2). The main learning target of the course is self-organisation in a multi-team Scrum setting. Therefore, we intended to minimize interference, but still had to evaluate how teams adopted the process. To achieve this balance, we implemented a threefold approach. Firstly, anonymous surveys helped to understand which elements of the process seemed most important or problematic for the students (Section 3). Secondly, tutors observed Scrum meetings with minimal interference to get a first-hand impression of Scrum adoption (Section 4). Finally, we collected and analyzed digital artifacts created by the students throughout the semester in order to define and detect anti-patterns that indicate potential misuse of the taught principles (Section 5).

While this setup allowed us to gather insights into Scrum adoption within the project, it still is at the level shown in Figure \ref{fig:coverage}(b), as automated analysis was not used in-situ. Hence, this paper marks an intermediate step towards the ideal scenario (Figure \ref{fig:coverage}(c)), in which surveys and tutor observations play a less dominant role but are more focused with the help of insights provided by automated tools, such as ScrumLint (Section 6). The paper concludes with learnings from applying the three observation techniques in our course setup (Section 7) and a brief overview of related approaches (Section 8).
 

%% file: course_setup.tex
\section{Course Setup}
\label{sec:course_setup}
The course under study is a final year undergraduate software engineering course, whose main idea is that multiple teams jointly develop a single system. All participants previously attended lectures teaching the fundamentals of software engineering, including core principles of agile development processes. Prior to the beginning of the course, an introduction exercise aims to familiarize participants with the programming environment (Ruby on Rails). To avoid introducing unnecessary overhead, we performed a short survey about prior experience (see Table~\ref{table:prevexp}). Only about one third  of the students had prior experience with the framework or web development in general. To refresh Scrum knowledge, a Lego exercise is performed in the second week of the course, prior to the project. Furthermore, lectures about other core elements of the course, such as Test-Driven Development, Behavior-Driven Development, Continuous Integration and Automated Deployment were given throughout the semester.

\begin{table}
	\centering
	\caption{Experience of students. 31 students answered the survey.}
	\label{table:prevexp}
	\begin{tabularx}{\columnwidth}{|X|r|}
	\hline
	\textbf{Previous experience with ...} & \textbf{\% \enquote{yes}} \\ \hline
work in software development projects (3+ people)? & 100.00\% \\ \hline
Scrum or other agile methodologies? & 61.29\% \\ \hline
the waterfall development model? & 25.81\% \\ \hline
Ruby / Ruby on Rails or web development? & 35.48\% \\ \hline
work as a paid software developer in a team? & 29.03\% \\ \hline
	\end{tabularx}
\end{table}

\subsection{Development Process}
\begin{figure}[!ht]
	\centering
	\includegraphics[width=\columnwidth]{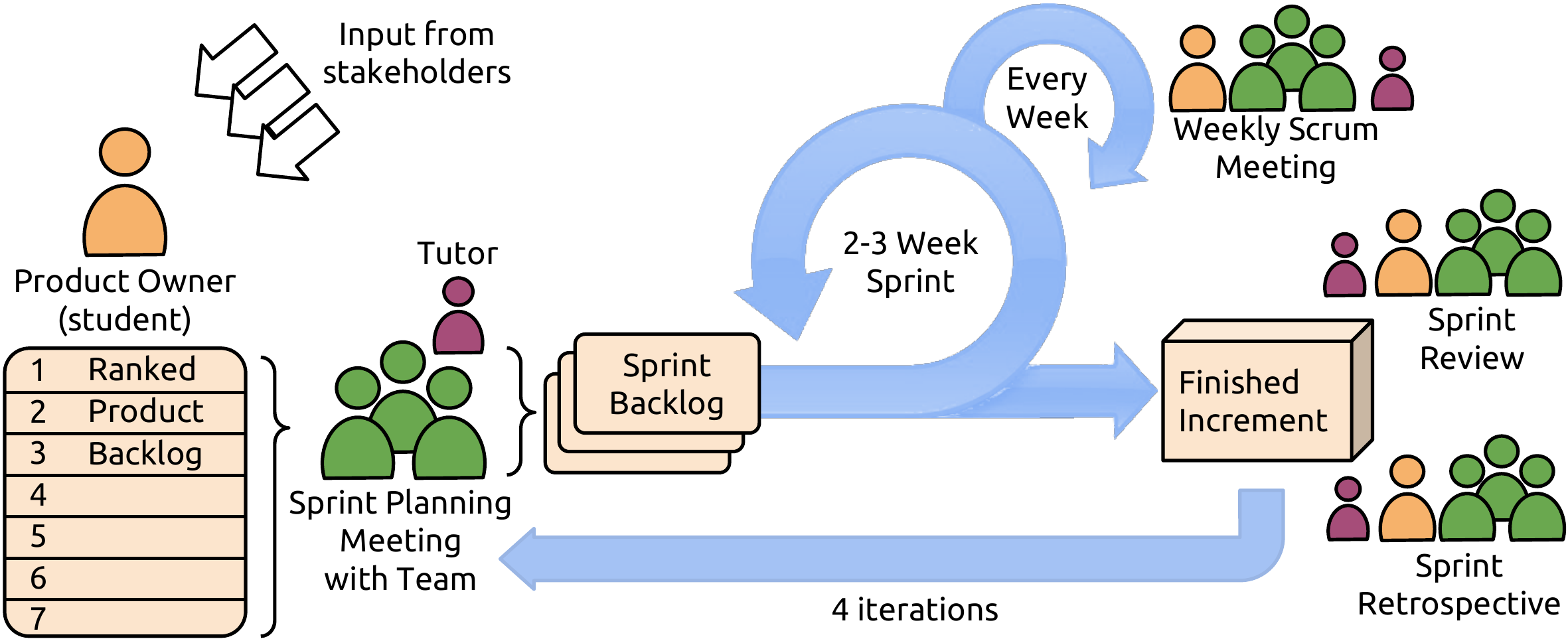}
	\caption{The modified Scrum process followed in the course Software Engineering II.}
	\label{fig:scrum_process}
\end{figure}

Within the project, a custom-tailored version of the Scrum process is used as a framework for development efforts (see Figure \ref{fig:scrum_process}). Based on the number of attendees, a varying number of teams is formed by the students, each consisting of 5 to 8 members. Within each team, one Product Owner (PO) and a Scrum Master (SM) is present, the remainder of the students act as developers. The course takes place in the winter term and allows for 12 weeks of active development, which are split into 4 sprints of three weeks duration. Within each week, students are required to spend 8 hours on the course, including participation in lectures.

For each sprint, a planning meeting, weekly Scrum meetings, a sprint review, and a retrospective meeting have to be organized by the students and performed in the presence of a tutor. Additional meetings, such as Scrum-of-Scrum meetings or gatherings of the POs were scheduled on demand. As student working time is supposed to be limited to 8 hours per week, daily standup meetings are replaced by weekly versions.

Two weeks prior to the project's start, the POs meet with the ``customer'' (i.e., a member of the teaching staff) and discuss initial requirements for the system. Under guidance by the teaching staff, they consequently fill the product backlog and decide which team is responsible for the implementation of which aspects of the system. Once these preparations are finished, the POs present the overall product vision and responsibilities to the other students. Subsequently, each team performs a sprint planning meeting to decide on the sprint backlog and identify potential dependencies to other teams. After two to three weeks of development, a sprint review and retrospection concludes the sprint. During sprints, the PO team presents intermediate progress to the customer in order to clarify existing and elicit new requirements.

\subsection{Infrastructure}
To support teamwork, an open-source development infrastructure was employed. The code base is hosted on Github\footnote{\url{https://www.github.com}}. Each team has their individual team branch and can add feature branches, if desired. To synchronize activities between teams, a development branch is present. Changes to this branch should be introduced as pull requests and be reviewed by other team members or even members of other teams, in case dependencies exist. A ``master'' branch contains a working version of the system and is only updated if all tests pass for the development branch and no features are marked as ``work in progress''. Test execution is performed automatically through the continuous integration platform Travis-CI\footnote{\url{https://www.travis-ci.com}}, which also triggers a deployment to Heroku\footnote{\url{https://www.heroku.com}} servers in case of successful tests. The deployed instances of the system are monitored by errbit\footnote{\url{https://github.com/errbit/errbit}} in order to notify developers in case of errors that occurred, e.g., during presentations to the customer.

For backlog maintenance, the project uses the standard Github issue tracker with an additional Scrum-view provided by Waffle.io\footnote{\url{https://www.waffle.io}}. By that, all teams have access to issues that other teams are working on and can leave comments, e.g. if they detect duplicates. For communication, a Slack\footnote{\url{https://www.slack.com}} team was created and dedicated mailing lists for the course and all teams were provided. Some teams fortunately had access to dedicated spaces as they also formed teams in their parallel bachelor's projects, which conclude the bachelor program at our institute. In that case, we encouraged them to also create physical Scrum boards, however, with the request of keeping them in sync with the digital systems.

\subsection{Discussion}
The course has been offered since the winter term of 2009/10, hence the discussed installment marks its sixth iteration. While the general course concept and development process remained constant over the years, the topics have changed from building small enterprise systems to more targeted applications with real end-customers, such as administration tools for university job offerings. In the 2014/15 iteration, a platform for event and room management was created\footnote{\url{https://github.com/hpi-swt2/event-und-raumplanung}}. Furthermore, the collaboration infrastructure was constantly updated to remain in sync with recent developments, e.g., by exchanging Subversion with Git or using newly available Software-as-a-Service solutions in favour of self-hosted systems. Thereby outages of systems could be reduced and existing knowledge of popular tools was utilized.

Regardless of the topic or the infrastructure, the student teams cited similar challenges in the evaluation surveys performed at the end of each course. Eliciting, structuring, and communicating requirements was deemed to be the most challenging task and had major implications on collaboration, both within and between teams. Other courses, that also aim to teach agile methods, such as the one performed by Mahnic \cite{Mahnic2010}, overcome these issues by letting the teaching team create the backlogs. As our evaluations also stipulated that self-organization was a main takeaway from the course and students noted that they ultimately got accustomed to the process during the project, we refrained from this option. Instead, we implemented the multi-tiered observation and evaluation process, which will be presented in the next sections.

%% file: surveys.tex
\section{Surveys}
Surveys are a time-tested approach to gaining insights into how the quality of performed work is perceived by team members. They allow gathering knowledge of parts of the workflow where no direct supervision is possible. In the previously presented course, all members of the development teams filled out regular surveys. These dealt with their satisfaction with Scrum and their implementation of Scrum practices over the span of the course. In particular, the surveys had the following aims:

\begin{itemize}
    \itemsep0em 
	\item Find topics that students have issues with, so that these areas could be addressed in a timely manner in the course.
	\item Find correlations between patterns of answers to questions, allowing insights into what areas of the Scrum process are significantly related. This can indicate which key components should be measured and improved in order to affect larger parts of the development process.
\end{itemize}

Surveys were conducted at the end of each development sprint and consisted of two parts:
\begin{itemize}
    \itemsep0em 
	\item \textbf{Primary Survey} Survey of ten questions tracking key aspects of agile practices, based on existing work.
	\item \textbf{Supplementary Survey} Survey tracking additional aspects, which were missing in the primary survey and were more specific to our course setup.
\end{itemize}

\subsection{Primary Survey}
The primary survey (S1) aimed at tracking students' satisfaction with different aspects of the Scrum method, such as effort estimation and backlog maintenance as well as the perceived level of cooperation with various team members. These were goals shared by previous work performed by Mahnic~\cite{Mahnic2010}.
Adapting this existing survey allowed a level of comparison to previous courses.
The primary survey consisted of 10 questions (see Table~\ref{table:primarysurvey} for an extract), answerable on a 5-point Likert scale with grades 1 through 5: \enquote{strong no} (1) to \enquote{strong yes} (5), with 3 being \enquote{neutral}.

While the original survey was adequate for the more controlled, restricted setup of Mahnic's course, modifications were made to the wording of questions to adapt them to our more self-organized context. For example, the concept of requirements was replaced with that of user stories and references to material provided by educators of the original course were removed.
Furthermore, some aspects of teamwork, specific to our course, e.g. the product vision of Product Owners, were not covered by the original survey, as students had no influence in those areas. To gain insights into these additional areas, a second survey was devised.

\begin{table}
	\centering
	\caption{Excerpt of questions from S1\protect\footnotemark.}
	\label{table:primarysurvey}
	\begin{tabularx}{\columnwidth}{|l|X|}
	\hline
	\textbf{\#} & \textbf{Question} \\ \hline
	1 & \textit{Clarity of requirements in the Product Backlog} Were the user stories in the backlog clear enough? Did the descriptions suffice to understand what the Product Owner really wanted? \\ \hline
	2 & \textit{Effort estimation} Were the estimates of required work (story points / man-hours) of user stories adequate / realistic? \\ \hline
	5 & \textit{Cooperation with the Scrum Master} Was the cooperation with the Scrum Master adequate / satisfactory? Did the Scrum Master contribute to the team's success?\\ \hline
	\end{tabularx}
\end{table}

\subsection{Supplementary Survey}
The supplementary survey (S2) was conducted with teams together with the primary survey starting at the end of the second Sprint.
It included questions about topics not covered by the primary survey, which had been problem areas in previous installments of the software engineering course: testing, cooperation between teams, the quality of the backlog, and the \enquote{Definition of Done} (see Table~\ref{table:supplementarysurvey}).
Furthermore, aspects of Scrum were included in the survey that were adapted to the specific context of our course: the quality of the product vision (a measure of how well the product owners of each of the different teams worked together) and the quality of Weekly Scrums (instead of Daily Scrums).
The supplementary survey consisted of six questions.
They were answerable on a scale of 1 (\enquote{very good}) to 6 \enquote{insufficient}, representing German school grades, which were intuitive to students of the course.

\footnotetext{Available at: \url{https://github.com/chrisma/icse16}.}

\begin{table}
\centering
\caption{Excerpt of questions from S2\protect\footnotemark[8].}
\label{table:supplementarysurvey}
	\begin{tabularx}{\columnwidth}{|l|X|}
	\hline
	\textbf{\#} & \textbf{Question} \\ \hline
	1 & How do you rate the \textit{quality of tests} (amount, coverage, relevance, effectiveness, regression, etc.)? \\ \hline
	2 & How was the \textit{cooperation / communication with other teams} (what is being done, when, why, blockers, merges, problems are being solved collectively, etc.)? \\ \hline
	3 & How do you rate the \textit{product vision of the POs} (is it clear what the product should accomplish, POs speak with a unified voice, etc.)? \\ \hline
	\end{tabularx}
\end{table}

\subsection{Limitations}
For all surveys presented here, no identification of individual subjects took place.
Subjects were completely anonymous within their teams and were not tracked over time, e.g. by pseudonyms.
We felt this was necessary to preserve the students' trust in the grading process of the course.
Furthermore, participation in the surveys was voluntary, and was not part of the course description when students signed up.
As such, individual subject tracking was eschewed in order to maximize willingness to participate in the surveys and minimize the effort needed by students to fill them out.

\subsection{Summary}
The surveys created an overview of how Scrum team members perceive the enacted process.
However, conducting regular surveys is time intensive for both students and teaching staff and provides little detail about the causes of problems.
Especially, no concrete link to development artifacts that could have been a point of contention, can be obtained to allow further research.
Furthermore, due to the nature of surveys, in order to allow meaningful comparisons, surveys cannot be iterated and adapted during a course.
To alleviate some of these problems and provide a more objective view, in comparison to self-assessments, tutors were employed.

%% file: tutor_scores.tex
\section{Tutor Scores}
Members of the teaching staff with knowledge of Scrum practices and experience employing them were present at the Sprint Review and Planning meetings of the development teams.
These tutors gave feedback and evaluated students.
As meetings should be as natural as possible and should take the form that the specific student team preferred, tutors had the role of passive observers in the background.
They only interfered in meetings when directly asked by students or when blatant violations of Scrum principles occurred, that would threaten the success of the meeting.
Otherwise, students received feedback at the end of meetings in an \enquote{I like, I wish} format (a variant of Cockburn's \enquote{Keep/Problem/Try}~\cite{cockburn06}).

\begin{figure}
	\centering
	\includegraphics[width=\columnwidth]{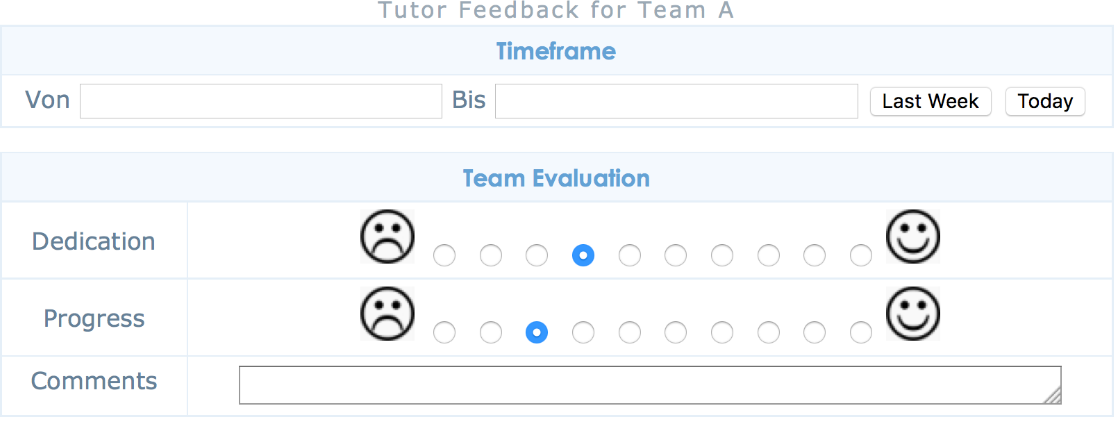}
	\caption{Online application for capturing tutor feedback. Team scores are automatically applied to all members, but can be changed on demand. Commentary fields are optional.}
	\label{fig:tutor_eval}
\end{figure}

The tutors rated the students on two aspects: their dedication to Scrum and associated methods as well as their progress in implementing these.
Teams that were very enthusiastic and embraced Scrum practices would score high on the \enquote{dedication} scale, whereas to be rated highly on the \enquote{progress} scale, a team would have to be efficient and deliberate in meetings and show understanding of Scrum practices and their adaptations.
Tutor scores were recorded in an online system which allowed team ratings on a score of 1 through 10 as well as comments on the team and individual students (see Figure \ref{fig:tutor_eval}).
Five meetings per team were evaluated, the first was the planning meeting for the first sprint, the last the sprint review of the fourth sprint.

The scores assigned by tutors represent quantified, more objective data on how well teams implemented Scrum.
As tutors were present at meetings, immediate, detailed feedback could be given to teams.
Tutors were able to apply their knowledge and experience with Scrum to students' situations and lend support with advice.
However, supporting teams with tutors is costly in terms of time and placing a member of the teaching staff into meetings can influence students' behavior~\cite{adair1984hawthorne}.
Because tutors had a common understanding of scoring standards a level of comparability between scores was supposed to be reached. At the same time, there were different opinions and attitudes towards Scrum and its application between tutors, which influenced assessments.
In order to reach the best possible comparability of scores, the same person would have had to attend every meeting.
Though this could be possible through strict scheduling of meetings, it does not scale well (e.g., a previous iteration of the lecture that included 13 teams \cite{kowark2011educational} would be impossible to handle by a single tutor in a reasonable manner) and is error-prone, e.g., in case of sickness.
Hence, we focused on a data-driven approach as the third pillar of Scrum assessment.

%% file: tool_analysis.tex
\section{Tool-supported analysis}
If the main goal of our software engineering course was to assess code and test quality, code coverage or \emph{Lint} code quality tools~\cite{Johnson1978} could be used to objectively assess whether students adhered to taught coding principles.
However, we aim at measuring how well teams implemented Scrum.

As such, it is not sufficient to restrict analyses to code, but all collaboration artifacts that reflect teamwork, should be included.
In order to find \enquote{anti-patterns}, instances where students had not followed agile principles, development artifacts, mainly user stories and commits into the software repository, were collected and analyzed.
These can give an insight into the work of students while they are working without supervision in their teams.
The 2014/15 software engineering course produced 379 user stories with 4707 revisions and 1802 commits featuring 26503 file changes.
As such, manually finding areas of improvement, where students deviated from agile practices, is cumbersome and scales poorly with the amount of participating students.
Automating this process with software can help gain new insights into the implementation of agile methods and best practices.
It provides a constantly available resource of feedback for students, an approach not unlike that of code quality tools, pointing out areas where the process could be improved (i.e. \enquote{process smells} instead of \enquote{code smells}~\cite{fowler99}).

\begin{figure}[!ht]
	\centering
	\includegraphics[width=0.75\columnwidth]{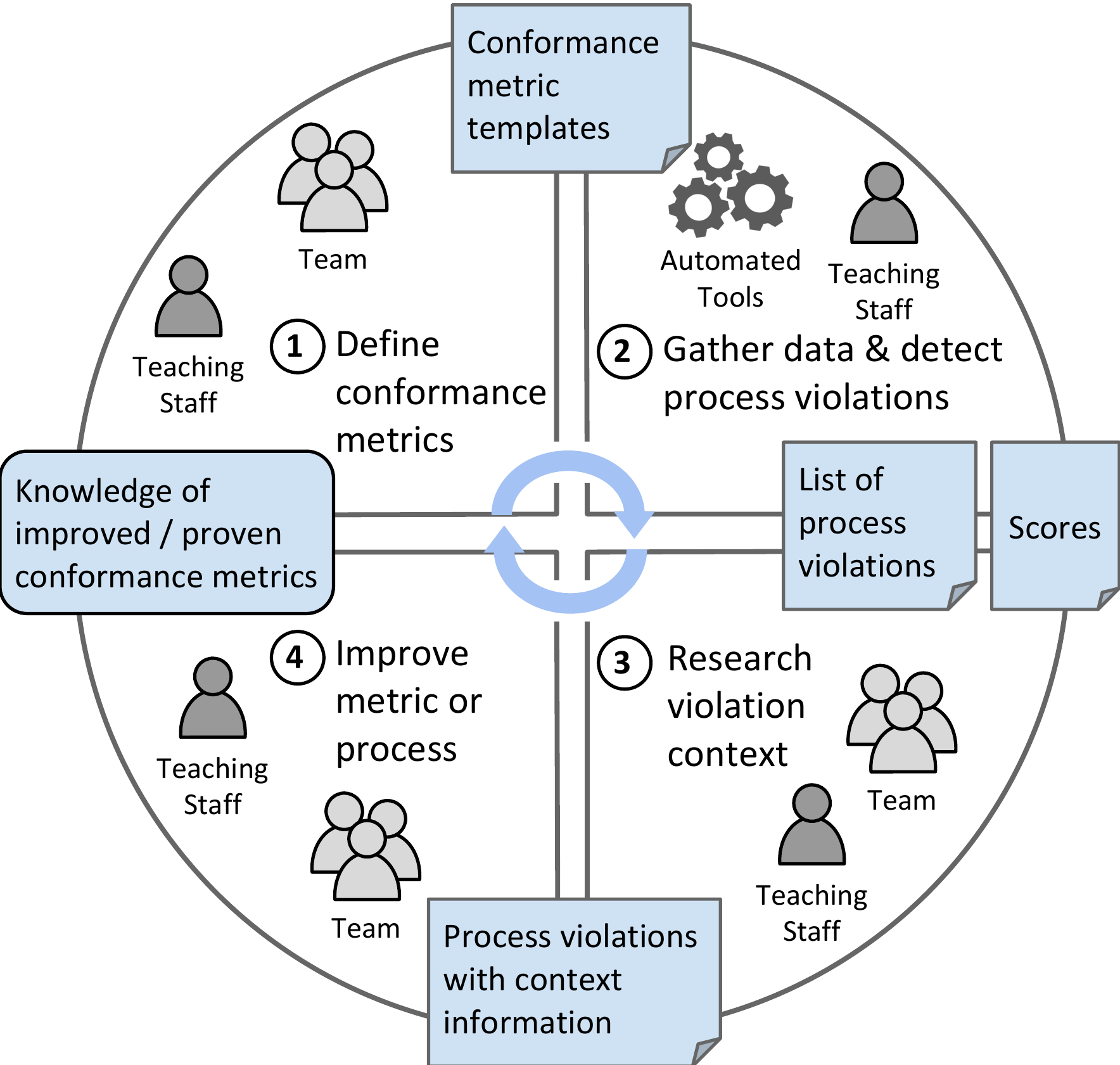}
	\caption{Overview of the conformance metric lifecycle. For each step of the lifecycle the involved parties as well as the in-and outputs are defined.}
	\label{fig:lifecycle}
\end{figure}

The main challenge with this approach is collecting and formalizing Scrum process violations.
We adapted Zazworka et al's.~\cite{zazworka2010developers} model of process nonconformance as a basis for detecting instances where processes were violated.
Figure~\ref{fig:lifecycle} describes the lifecycle of a \enquote{conformance metric}, which includes information about the agile practices that are measured, as well as the specifics of how to measure and evaluate deviations.
The lifecycle includes four steps, which are described in the following subsections.

\subsection{Define Conformance Metrics}
The first step in defining a conformance metric is to gain a common understanding of the practice that should be executed and measured.
This involves both the teaching staff, who have knowledge and experience in agile development, as well as the team, who might have personal preferences or different understandings.
Agile methodologies such as XP or Scrum advise a multitude of practices, e.g. \enquote{all code must have unit tests}. Sletholt et al. mention 35 main ones~\cite{Sletholt2012}, which can serve as a starting point to select practices that are applicable in the context of a certain project.
If a process is considered relevant enough to be measured and a common understanding of its details is found, this knowledge should be recorded in the form of a process conformance template\footnotemark[8].
	
\subsection{Gather Data and Detect Process Violations}
After an initial set of metrics is created, the required development data can be collected and, using the recorded information, potential violations of the process can be detected in an automated fashion.
The output of this step is a list of indicators for possible process violations, grouped by teams and iteration, as well as a score that indicates their severity. Details of the developed tool and its implementation are described in Section~\ref{sec:scrumlint}.

\subsection{Research Violation Context}
Once detected, additional information about the identified violations should be gathered. This can include researching details of the offending development artifacts, e.g. reviewing the change history of a user story or talking to the involved developers directly to gain an insight into what happened.
This step is necessary in order to gauge the quality of violations. These may only be symptoms of deeper problems, which should be tackled in the following improvement step. Furthermore, violations may not actually point to problems but can be false positives. This can be the case when queries are not refined enough, some special cases were forgotten, or the development context was not considered.
The output of this step should be a list of violations with additional information indicating priority and cause.

\subsection{Improve Metric or Process}
In the last step, measures on how to improve the metrics themselves as well as the overlap between defined and executed practice should be decided and implemented. If metrics return false positives, these should be eliminated by fine-tuning the way the metric detects process violations.
Two general courses of action lend themselves to reducing the amount of real violations, i.e. true positives, in the future: modifying the metric or improving the executed process.
In the first case, if the found violations are not considered harmful to development, there is little benefit in continuing to classify them as violations. The defined process can be adjusted to better fit the executed practices, resulting in less detected process violations.
Alternatively, process conformance can be improved by modifying the executed day-to-day process, making sure the defined process is followed. This can be achieved, for example, by directly reminding teams about the defined process and its benefits.

%% file: scrumlint.tex
\section{ScrumLint}
\label{sec:scrumlint}
\emph{ScrumLint} is the implementation of the automated \enquote{linting} approach that was developed for the course.
It gives feedback to students and educators on the level of conformance to Scrum practices in a team.
The main focus is implementing the \enquote{gather data \& detect process violations}

\begin{wrapfigure}{r}{0.45\columnwidth}
    \begin{center}
        \includegraphics[width=0.38\columnwidth]{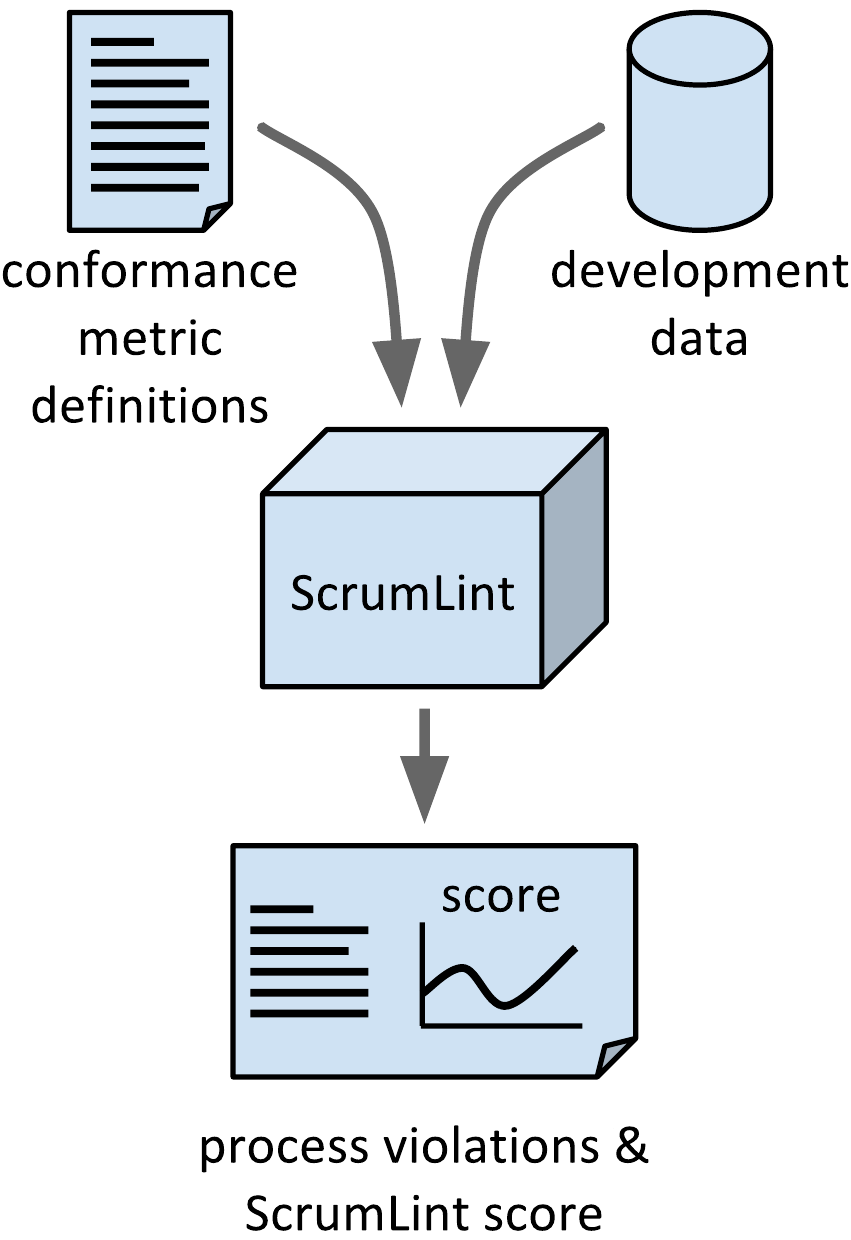}
    \end{center}
    \caption{In- and outputs of ScrumLint.}
    \label{fig:inandoutputs}
\end{wrapfigure}

\noindent
step of the conformance metric lifecycle.
ScrumLint is open-source software and is publicly available on Github\footnote{\url{https://github.com/chrisma/ScrumLint}} under the MIT license.
Figure~\ref{fig:inandoutputs} shows the in- and outputs of the application. Inputs are development data collected during the course and the conformance metrics to run on the data. Using these two sources of information, ScrumLint produces a list of detailed violations of the defined practices for each sprint and team.

\subsection{Data Collection}
\begin{figure}[!ht]
	\centering
	\includegraphics[width=\columnwidth]{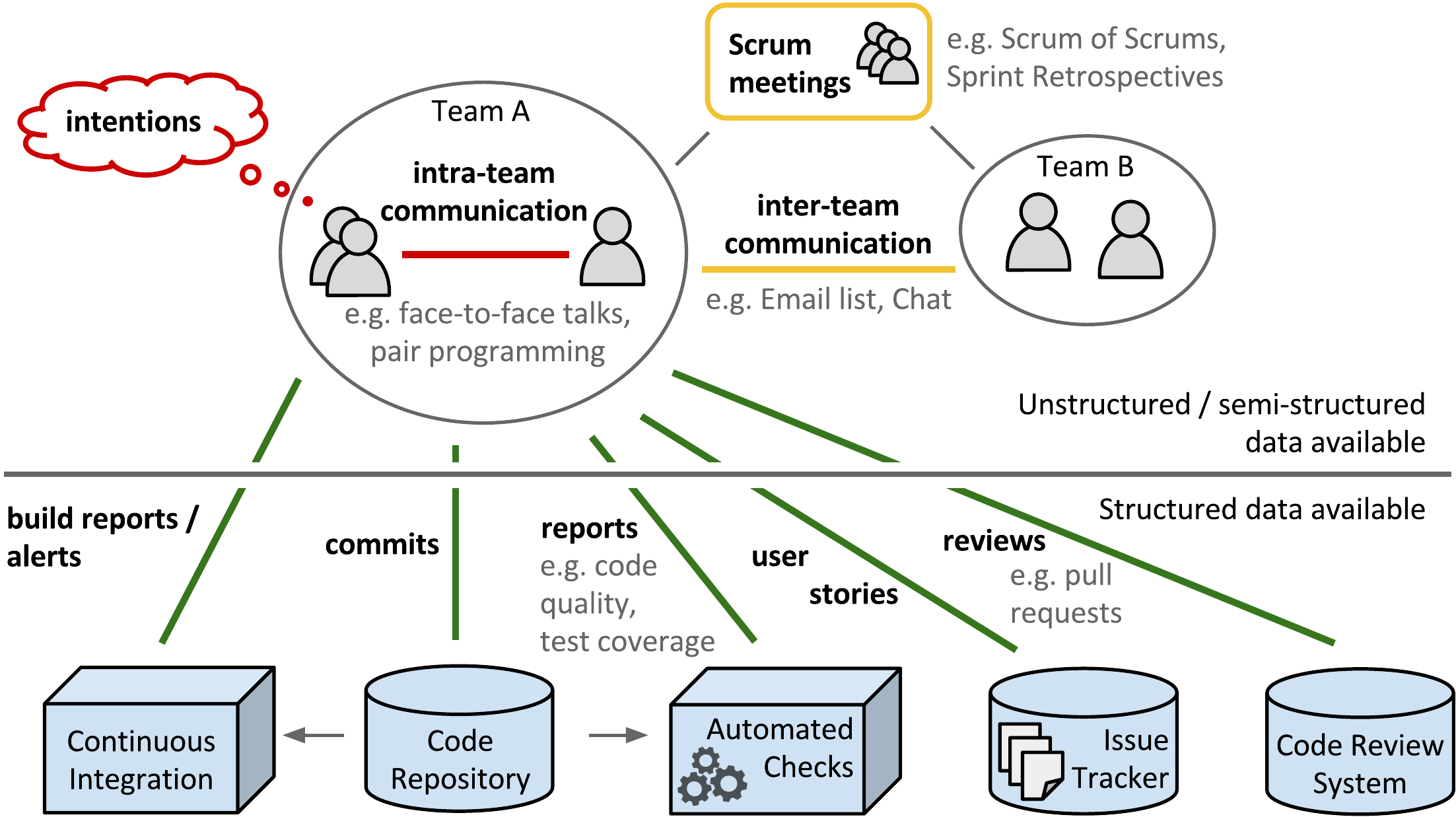}
	\caption{Examples of data sources. Structured data (bottom) is easier to collect (green) than more unstructured data (yellow) or data that needs to be collected completely manually (red).}
	\label{fig:datasources}
\end{figure}

Ideally, as much data as possible is collected automatically from already existing development artifacts and data sources in order to minimize collection effort for team members and teaching staff. 
This avoids introducing additional constraints (e.g., prescribed usage of custom IDEs with pre-installed monitoring tools) and documentation duties, which go against the \enquote{self-organizing} spirit of Scrum and agile development.
Devedzic et al. claim \enquote{greater involvement and greater level of interaction}~\cite{Devedzic2011} as a result of self-organizing teams.
Gathering data from a development process can be more or less complex, depending on how structured the data is which is to be collected. Figure~\ref{fig:datasources} gives examples of different data sources.
Structured data, such as a commit to a version control system, is less complex to collect and analyze, compared to semi-structured data such as an E-Mail, which includes large amounts of free text. 
For the first version of ScrumLint we focused mainly on source code management and issue tracker data, as these represented the majority of data that students produced regularly.

The primary data source for development information was Github's git repository, as the main part of the software, the code, was hosted there.
As Github features a comprehensive JSON API\footnote{\url{https://developer.github.com/v3/}} which exposes every piece of information that is available within the system, all the data associated with a git repository could be retrieved. 
This includes, in particular, information associated with commits, i.e. committer, date and time, commit message and changed files.
Github also includes the filepaths of files that were changed and reports statistics about each commit, such as amount of changed and added lines.
Data collected from source code management systems enables measuring the rate and details of which parts of the source code changed and who made the modifications.
It needs to be noted that skilled developers could alter this data and thereby reduce its validity \cite{Bird:2009aa}.

Github's issue tracker was used to store and manage user stories created by POs.
Although the issue tracker is not explicitly designed for this purpose, some limitations were worked around, e.g. adding priorities and effort estimations to the titles of tickets\footnote{\url{https://github.com/hpi-swt2/event-und-raumplanung/issues/263}}.
Through the Github API, information regarding the tickets / user stories, including  titles and content, as well as all changes and their timestamps, e.g. \emph{assigned to Team A at 12:30 AM on Jan 14, 2015}, could be retrieved.
Collecting data on user stories enables, for example, measurements of how fast user stories in the Sprint Backlog were considered done and were closed.

Collected development data is written to a database.
Due to the strongly-connected and relationship-focused nature of the data, e.g. every commit has an author, every author has a team, every file or ticket change is related to a file or ticket, a graph database\footnote{Neo4j (\url{http://neo4j.com/}) was used.} was employed.
This allows easily adding data from additional data sources without having to adopt a generic database scheme from the beginning on or having to adapt a specific database scheme when changes occur.
In particular, semi-structured data can easily be written to the database.
Labels and connections can be added afterwards as information from queried data sources is added.

\subsection{Data Analyses}

Once data is collected, conformance metrics are used to analyze it.
These include a description of the measured agile principle as well as a database query that is able to extract instances of process violations (see Table~\ref{table:conformancemetric}). Queries are defined using Neo4j's Cypher query language and can be easily adapted through an administrative interface.

\begin{table}
	\centering
	\caption{Example of a conformance metric.}
	\label{table:conformancemetric}
	\begin{tabularx}{\columnwidth}{|X|}
	\hline
	\textit{Name:} The Neverending Story \\ \hline
	\textit{Synopsis:} User stories that were in multiple sprint backlogs. \\ \hline
	\textit{Category:} Backlog Maintenance \\ \hline
	\textit{Severity:} High \\ \hline
	\textit{Effort}: Low \\ \hline
	\textit{Data source:} User story tracker \\ \hline
	\textit{Rating function:} $max(0, 100-(\frac{\#violations}{\#totalUS}*100*AvgInSprints*weight))$, $\#violations$ = amount of user stories in more than  $threshold_{amount}$ sprints, $\#totalUS$ = total amount of user stories in the sprint backlog, $AvgInSprints$ = average amount of sprint backlogs the violations were in. \\ \hline
	\textit{Description:} Ideally, a sprint backlog contains exactly as many user stories as the team can complete in the iteration~\cite{Schwaber2011}, meaning that at the end of the sprint all user stories in the sprint backlog are finished. This ensures the ability to plan the software's development and enables teams to build on the finished functionality in the next sprint. However, sometimes, at the end of the sprint not all stories conform to the \enquote{Definition of Done}~\cite{Kniberg2007}. These user stories are then carried over to the next sprint, if the product owner still considers them a priority. A story that spans multiple sprints can be a blocker for other teams that depend on it. This metric identifies user stories that were assigned to the sprint backlog of multiple sprints, with or without commits referencing it. \\ \hline
	\end{tabularx}
\end{table}

Currently ScrumLint contains ten conformance metrics in the categories \enquote{teaching scores}, \enquote{developer productivity}, \enquote{Backlog maintenance} and \enquote{XP practices}, that were developed for the course.
These include more general measurements, e.g. the \enquote{bus number}~\cite{ricca2011difficulty}, as well as measurements specific to the context of our course, e.g. the percentage of commits in the last 30 minutes before sprint deadline.

\subsection{Displaying Results}
\begin{figure}[!ht]
	\centering
	\includegraphics[width=\columnwidth]{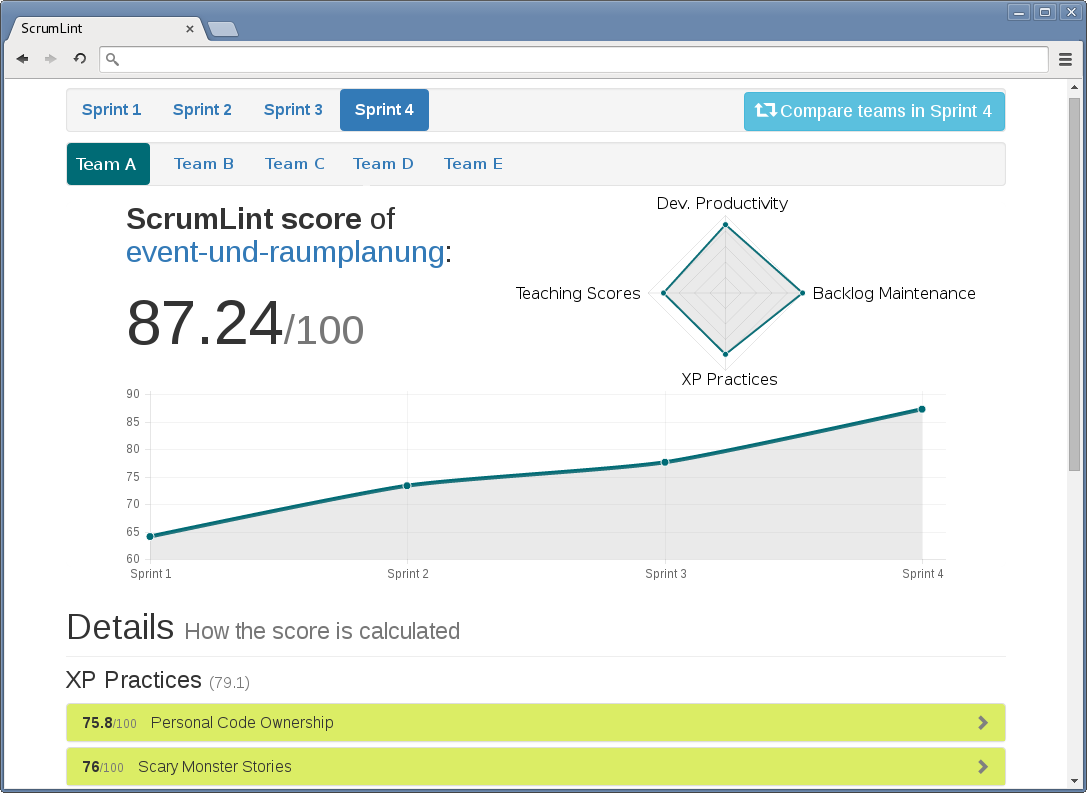}
	\caption{Screenshot of ScrumLint showing the score for Team A in Sprint 4. Details of found process violations are available below the fold.}
	\label{fig:teamoverview}
\end{figure}

The results of analyses are presented to users as a webpage, thereby allowing access without installing additional software. 
In ScrumLint, a score, reflecting the severity of violations, is assigned to each metric.
These individual scores are combined to form an overall ScrumLint score for a team in an iteration.
A screenshot of ScrumLint showing a team's score is given in Figure~\ref{fig:teamoverview}.

\begin{figure}[!ht]
	\centering
	\includegraphics[width=0.7\columnwidth]{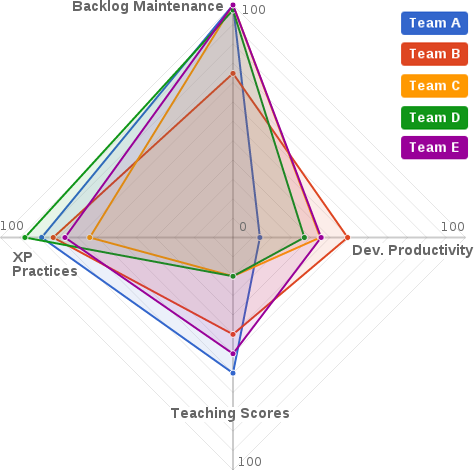}
	\caption{Radar chart comparing teams by categories of metrics.}
	\label{fig:radar}
\end{figure}

We chose a range of 0 to 100 for the score, where 100 indicates that no violations were found while 0 indicates that the defined practices were not followed at all.
However, any scale that is intuitive to users can be used.
The development of scores and comparisons between teams are visualised using line and radar charts (see Figures~\ref{fig:teamoverview},~\ref{fig:radar}).

\begin{figure}[!ht]
	\centering
	\includegraphics[width=\columnwidth]{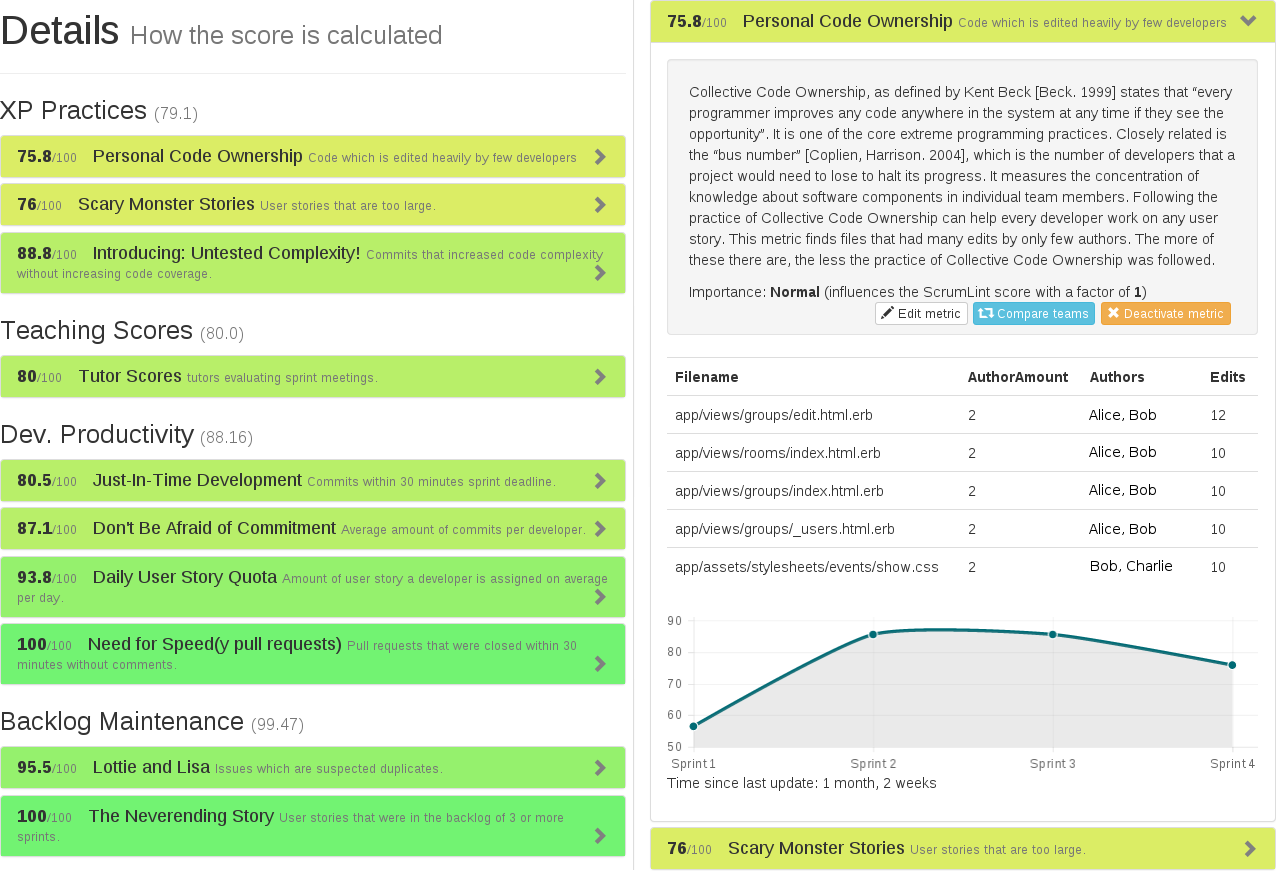}
	\caption{ScrumLint screenshots. List of conformance metrics ordered by their scores (left) and a specific metric's details expanded (right).}
	\label{fig:teamdetails}
\end{figure}

The \emph{ScrumLint score} can be used to prioritize and order metrics and associated violations.
Figure~\ref{fig:teamdetails} shows the individual scores of metrics for a team in an iteration.
The lower the score, the more violations were found and the more likely it is that there is an actual problem in the way a team works.
The details returned by metrics, i.e. the development artifacts that were identified to be in violation of Scrum practices, can be used to research the context of the violation.

This can consequently help to improve the executed process or the metric itself.
For example, if a commit is found to significantly decrease code coverage, violating the \enquote{test first} principle, the changes can be viewed in detail on the commit's Github page.
Based on the additional information, informed decisions can be made, such as talking to the the developer that made the commit or changing the testing setup.

\subsection{Summary}
ScrumLint allows executing and visualizing a collection of defined agile conformance metrics for a given project.
It explicitly takes into account agile concepts such as user stories, working in development teams, and iterations.
Results are aggregated into scores, split by team, iteration, and category that allow a compact representation of the conformance to agile practices.
Representing results grouped by iterations allows comparing conformance to Scrum practices over time and integrating ScrumLint into the existing Scrum cycle, e.g. using it in Sprint Retrospectives.
Existing workflows do not need to be changed significantly as most of the currently implemented metrics rely on automatically collected data from existing development artifacts.
ScrumLint can alleviate the need to manually analyze large amounts of development data, allowing the focus on those that might pose a problem.
Violations of agile practices can be tracked down to the actual development artifact, allowing discussions on the basis of concrete data.

As conformance metrics are the basis of ScrumLint, their quality is mainly responsible for the quality of overall results.
However, there is little research yet on what constitutes best practices for agile metrics.
Zazworka et al. state that the \enquote{biggest challenge was to find definitions for the XP practices that contained enough detail}~\cite{zazworka2010developers}.
However, using ScrumLint and a relatively small amount of metrics (10 overall), we were already able to extract areas of improvement in the Scrum workflow of student teams for all iterations.

%% file: learnings.tex
\section{Learnings} 
Using the three presented sources of data on how well teams implemented and embraced Scrum (surveys, tutors, and software) it is possible to gain a more complete picture of how development teams work.
Combining these approaches allows deeper insights than relying on only a subset of these sources.

\subsection{Surveys}
The results of the primary survey in Table~\ref{table:survey1} show that the null hypothesis of students having a neutral attitude towards Scrum, was rejected in the vast majority of questions: 8/10 after Sprint 1, 7/10 after Sprint 2, 8/10 after Sprint 3, 9/10 after Sprint 4.
All of these questions show average ratings higher than 3, indicating a positive attitude towards Scrum and its practices.
We can therefore lend some credibility to the hypothesis that students on average had a positive attitude towards the use of the Scrum methodology in our course.
In particular, the average rating for question 10, asking whether students were satisfied with Scrum and would recommend it, only sank below 4 (\enquote{yes}) once, after Sprint 2 (average rating of 3.82).
These findings of students having a positive attitude towards hands-on learning of Scrum are in line with previous studies~\cite{Mahnic2010, Melnik2005}.
Using the answers to the regular surveys we were furthermore able to find topics that students had issues with and could respond to them in a timely manner in meetings and lectures.

\begin{table*}[htb]
\centering
\begin{tabular}{|l|lllllllll|}
  \hline
\textbf{Survey topic} & \textbf{1.} & \textbf{2.} & \textbf{3.} & \textbf{4.} & \textbf{5.} & \textbf{6.} & \textbf{7.} & \textbf{8.} & \textbf{9.} \\ \hline
1. Backlog clarity &  &  &  &  &  &  &  &  &  \\
2. Estimation &  0.04  &  &  &  &  &  &  &  &  \\
3. Backlog Maintenance &  0.14  &  0.14  &  &  &  &  &  &  &  \\
4. Administrative Work & -0.31*** &  0.18*  & -0.18*  &  &  &  &  &  &  \\
5. SM collaboration &  0.00  &  0.12  &  0.49*** & -0.12  &  &  &  &  &  \\
6. PO collaboration &  0.49*** &  0.08  &  0.31*** & -0.23**  &  0.28*** &  &  &  &  \\
7. Team collaboration &  0.34*** & -0.10  &  0.17*  & -0.21*  &  0.16*  &  0.32*** &  &  &  \\
8. Overall Workload &  0.23**  &  0.22**  &  0.14  & -0.11  &  0.28*** &  0.24**  &  0.11  &  &  \\
9. Satisfaction Work &  0.30*** &  0.32*** &  0.14  & -0.03  &  0.15  &  0.39*** &  0.30*** &  0.27*** &  \\
10. Satisfaction Scrum &  0.06  &  0.09  &  0.26**  & -0.11  &  0.24**  &  0.23**  &  0.20*  &  0.09  &  0.19*  \\ \hline
\end{tabular}
\caption{Correlations between answers to questions of S1. The topics of questions are abbreviated. Correlation significance: ***<0.001, **<0.01, *<0.05.}
\label{table:correlation1}
\end{table*}

Mahnic reports that no significant correlations between the satisfaction with Scrum (q10) and other questions could be found~\cite{Mahnic2010}.
In our repeat of the survey (S1), however, we found significant positive correlations in answers to this question with all questions regarding collaboration with Scrum roles (see Table~\ref{table:correlation1}): collaboration with the SM (q5), collaboration with the PO (q6) and collaboration with the team (q7).
Similarly, in S2, answers to the question regarding collaboration with other teams showed correlation with satisfaction with Scrum in S1 (r=0.45).
While these results are somewhat expected as Scrum is focused on effective communication, the strongest correlation with the question regarding satisfaction with Scrum was with the backlog maintenance (q3 in S1, r=0.26).
This indicates that the satisfaction with Scrum depends on knowing how to effectively deal with user stories.
Furthermore, in S2, the perceived quality of tests (q1) also showed positive correlation (r=0.24) with the satisfaction with Scrum.
These results suggest that the satisfaction with Scrum of student teams not only depends on the perceived quality of collaboration, which is hard to determine automatically, but also depends on the perceived quality of development artifacts, such as user stories and tests.
Based on these findings we theorize that analysing and improving especially the quality of these development artifacts can help improve the overall Scrum process.

\begin{table*}[ht]
\centering
\begin{tabular}{|@{}l|lll|lll|lll|lll|l@{}|}
\hline
 & \multicolumn{3}{c|}{\textbf{Sprint 1}} & \multicolumn{3}{c|}{\textbf{Sprint 2}} & \multicolumn{3}{c|}{\textbf{Sprint 3}} & \multicolumn{3}{c|}{\textbf{Sprint 4}} & \multicolumn{1}{c|}{} \\ \hline
\multicolumn{1}{|c|}{\#} & \multicolumn{1}{c}{Mean} & \multicolumn{1}{c}{\begin{tabular}[c]{@{}c@{}}SD\end{tabular}} & \multicolumn{1}{c|}{\begin{tabular}[c]{@{}c@{}}TTest\end{tabular}} & \multicolumn{1}{c}{Mean} & \multicolumn{1}{c}{\begin{tabular}[c]{@{}c@{}}SD\end{tabular}} & \multicolumn{1}{c|}{\begin{tabular}[c]{@{}c@{}}TTest\end{tabular}} & \multicolumn{1}{c}{Mean} & \multicolumn{1}{c}{\begin{tabular}[c]{@{}c@{}}SD\end{tabular}} & \multicolumn{1}{c|}{\begin{tabular}[c]{@{}c@{}}TTest\end{tabular}} & \multicolumn{1}{c}{Mean} & \multicolumn{1}{c}{\begin{tabular}[c]{@{}c@{}}SD\end{tabular}} & \multicolumn{1}{c|}{\begin{tabular}[c]{@{}c@{}}TTest\end{tabular}} & \multicolumn{1}{c|}{\begin{tabular}[c]{@{}c@{}}ANOVA\end{tabular}} \\ \hline
1 & 3.56 & 0.84 & 3.79*** & 3.81 & 0.83 & 5.39*** & 3.97 & 0.51 & 11.38*** & 4.00 & 0.57 & 9.96*** & 1.47 \\
2 & 2.97 & 0.93 & -0.19 & 2.76 & 1.13 & -1.19 & 3.16 & 0.81 & 1.16 & 3.61 & 0.86 & 4.03*** & 1.21 \\
3 & 3.28 & 1.05 & 1.51 & 3.74 & 1.12 & 3.67*** & 3.91 & 0.73 & 7.43*** & 4.13 & 0.75 & 8.47*** & 6.52** \\
4\up{1} & 3.50 & 0.95 & 2.98** & 3.10 & 1.14 & 0.47 & 3.06 & 0.82 & 0.42 & 3.20 & 0.93 & 1.27 & 1.49 \\
5 & 4.18 & 0.73 & 9.34*** & 4.22 & 0.75 & 9.18*** & 4.26 & 0.60 & 12.26*** & 4.33 & 0.69 & 11.07*** & 0.43 \\
6 & 3.78 & 0.83 & 5.31*** & 4.13 & 0.75 & 8.47*** & 4.04 & 0.81 & 7.68*** & 4.26 & 0.73 & 9.61** & 1.22 \\
7 & 3.63 & 0.91 & 4.09*** & 3.91 & 1.01 & 5.16*** & 4.01 & 0.54 & 11.25*** & 4.07 & 0.64 & 9.85*** & 2.44" \\
8 & 3.68 & 0.67 & 6.14*** & 3.26 & 0.96 & 1.60 & 3.42 & 0.64 & 4.04*** & 3.49 & 0.84 & 3.47** & 1.53 \\
9 & 3.41 & 0.90 & 2.75** & 3.40 & 1.12 & 2.12* & 3.40 & 0.89 & 2.74** & 3.94 & 0.86 & 6.59*** & 1.78 \\
10 & 4.14 & 0.67 & 10.25*** & 3.82 & 0.87 & 5.52*** & 4.07 & 0.47 & 13.99*** & 4.10 & 0.61 & 10.84*** & 1.44 \\ \hline
\end{tabular}
\caption{Data from S1, answers over the whole body of students. \textit{ANOVA} shows f-value of an analysis of variance on the team means for one factor (sprint). Significance: ***<0.001, **<0.01, *<0.05, "<0.15. \up{1}indicates items with inverse item coding.}
\label{table:survey1}
\end{table*}

\subsection{Perceptions of Tutors}
\begin{figure}[!ht]
	\centering
	\includegraphics[width=\columnwidth]{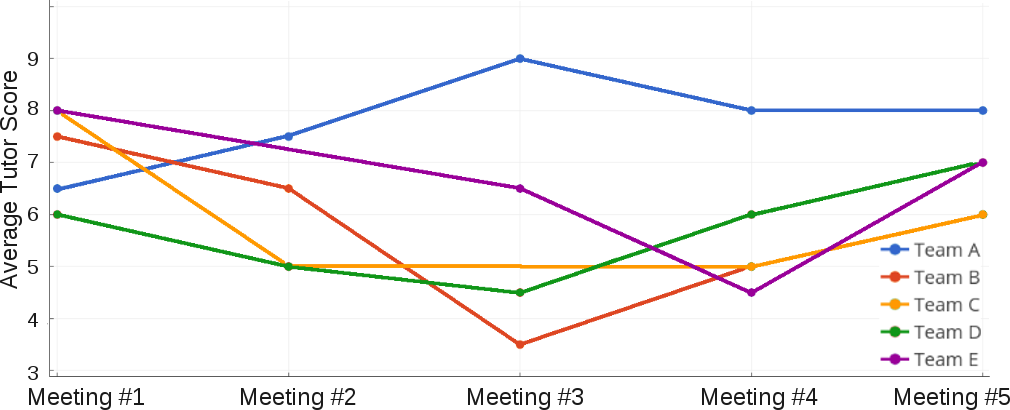}
	\caption{Average tutor scores by team and meeting. Gaps in the data were connected for a better overview.}
	\label{fig:tutorscores}
\end{figure}

While surveys provide self-assessments of students, tutors provide a view from the outside that is vital for judging Scrum adoption in a team.
These tutor scores are important in allowing tutors to see the recent development of a team at a glance and knowing which teams need more attention and support.
For example, Team A scored much higher than the rest of the teams and was thought of by tutors to have a very good working atmosphere, while Team B received below average scores in the middle of the course (see Figure~\ref{fig:tutorscores}).
This development is not as noticeable in the corresponding surveys, further highlighting the need for a multi-tiered approach.

Scores assigned by tutors in Scrum meetings reflect not only conformance of a team to the Scrum process but are also influenced by how the team adapted the process and worked as a whole.
For example, students' self-assessments might be low after an overlong meeting with little immediate results.
However, a tutor might rate the Sprint Review and Retrospective of this team highly, because the students  tried new techniques, reflected on their problems and proposed solutions.
With more knowledge of the Scrum process, tutors are able to determine whether students have learned and progressed, both in Scrum practices as well as in organizational aspects.
\newline
\subsection{ScrumLint}
\begin{figure}[!ht]
	\centering
    \includegraphics[width=\columnwidth]{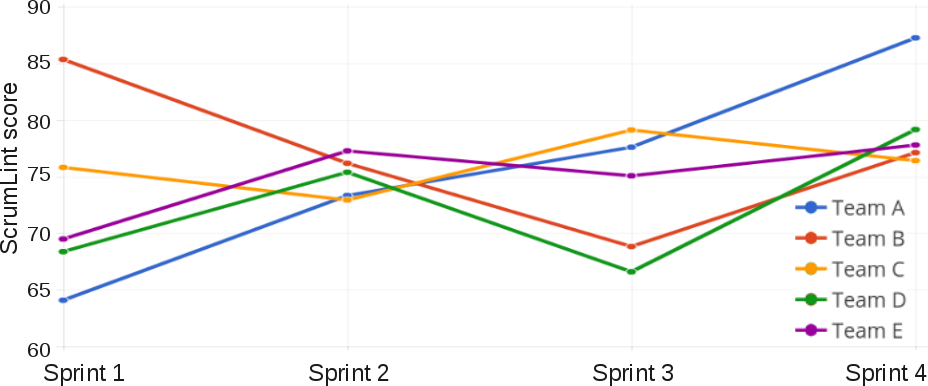}
    \caption{Graph of ScrumLint scores by team.}
    \label{fig:slscores}
\end{figure}

In the described course ScrumLint was able to detect process violations for every team in every sprint.
No team received a perfect overall ScrumLint score of 100 for an iteration; ScrumLint was able to highlight artifacts that needed extra attention for every sprint and team.
The highest score of 87.2 was assigned to Team A in the last sprint.
Figure~\ref{fig:slscores} shows the development of ScrumLint scores of teams over sprints.
It is important to note here that ScrumLint was not available to students during the course and did not influence the project.

The results of ScrumLint are different in nature to those collected from surveys or ratings by tutors.
In an educational context, ratings of students and teams reflect the learning progress in applying Scrum, which can involve deviating from the defined processes.
It is acceptable for students to \enquote{violate} the process if they learn from their experience and gain insights into why that Scrum practice was defined in the first place.
As such, ScrumLint is not a replacement for surveys or tutor assessments of teams.
Instead, it acts as an additional source of information for developers and teaching staff.
It can help in checking the perception of the state of the project against the actual, measured state of the development artifacts and give insights into what areas of the Scrum process could be improved in the future.

%% file: related_work.tex
\section{Related Work}
With the widespread adoption of agile methods in software engineering education, assessment is a frequent topic of related research. A proven choice, which was, for example, applied by Kropp and Meier \cite{Kropp2013}, is to perform post-hoc oral or written exams to determine whether the students obtained the desired knowledge, and judge the project outcome and presentations to assess success. Mahnic, on the other hand, used a series of surveys to determine whether the learning targets of his course were met \cite{Mahnic2010}, as well as the \emph{Earned Value method} (EVM) project management technique and its associated performance indexes in addition to Scrum burndown charts as tools to gain insight into project performance. Igaki et al. use an approach called ticket driven development (TiDD) to provide quantitative measurements about different aspects of the development process \cite{Igaki2014}. While it provides substantial insights into the inner workings of the formed Scrum teams, ticket creation introduces additional overhead and reduces the freedom teams have in adapting Scrum to their particular needs. With the combination of ScrumLint, tutor observations, and focused surveys, we aim to maintain this freedom while still creating comparable insights into the process.

Another stream of research focuses on measuring process conformance. Johnson et al. employed Hackystat-UH in order to perform a variety of fine-grained analyses of development behaviour \cite{Johnson04}. As the system relied on sensors within the students' IDE, freedom of tool choice was limited. Zazworka et al. defined violations of XP practices and measured them in a comprehensive study \cite{zazworka2010developers}. Their main challenge, however, was to find solid definitions of what constitutes a violation in the first place, and adapt it to the specific setup and teams. This is an issue that we aim to tackle with the flexible definition of conformance metrics in ScrumLint.

%% file: conclusion.tex
\section{Conclusions}
The combination of traditional surveys, tutor observations, and initial testing of automated process analysis allowed us to gain deeper insights into Scrum adoption within the presented course. Based on the general problem domains identified through the surveys, tutors were able to focus their feedback during meetings. Automated analysis post-hoc revealed evidence for certain adoption problems. We therefore conclude that the presented approach is viable for the setup of the course, but can also be adopted to other setups by refocusing the surveys and modifying ScrumLint metrics to cater to the peculiarities of the respective course. 

In future work, we will perform the course again, this time using ScrumLint in-situ. In the first iteration, the tool will be given to the tutors as an aid for preparing their meetings with the teams. In consecutive iterations, it will be interesting to see how student teams make use of such tools. While our hope is that they use them to question their process implementation and provide feedback about the significance of metrics, it is possible that they identify weak spots and simply try to avoid generating negative scores, e.g., through automation of tasks. In that case, however, it will be again interesting to see which influences such ``cheats'' have on the process and whether or not they accidentally improve Scrum implementation. Despite these future challenges, we see great potential in a combination of the presented observation techniques and think that they could reduce workload for the teaching staff while allowing to maintain or even increase process visibility in classroom projects.